\documentclass[11pt,a4paper]{article}

\usepackage{jheppub}

\usepackage{amsmath,latexsym,amssymb,tables,psfig,bbm,graphicx,eufrak}
\usepackage{macros,slashed}

\input eqalign.sty

\title{Short distance singularities and automatic O($a$) improvement: the cases of the chiral condensate
and the topological susceptibility}

\author[a,b]{Krzysztof Cichy}
\author[a,c]{Elena Garcia-Ramos}
\author[a]{Karl Jansen}
\affiliation[a]{NIC, DESY, Platanenallee 6, 15738 Zeuthen, Germany}
\affiliation[b]{Adam Mickiewicz University, Faculty of Physics,
Umultowska 85, 61-614 Poznan, Poland}
\affiliation[c]{Humboldt Universit\"at zu Berlin,  Newtonstrasse 15, 12489 Berlin,
Germany}

\preprint{DESY 14-192, HU-EP-14/42, SFB/CPP-14-80}

\emailAdd{krzysztof.cichy@desy.de}
\emailAdd{elena.garcia.ramos@desy.de}
\emailAdd{karl.jansen@desy.de}

\abstract{
Short-distance singularities in lattice correlators 
can modify their Symanzik expansion by leading to additional O($a$) lattice artifacts.
At the example of the chiral condensate
and the topological susceptibility, we show how to account
for these lattice artifacts for Wilson twisted mass fermions
and show that the property of automatic O($a$) improvement 
is preserved at maximal twist.
\begin{center}
\vspace*{1cm}
\includegraphics
[width=0.2\textwidth,angle=0]
{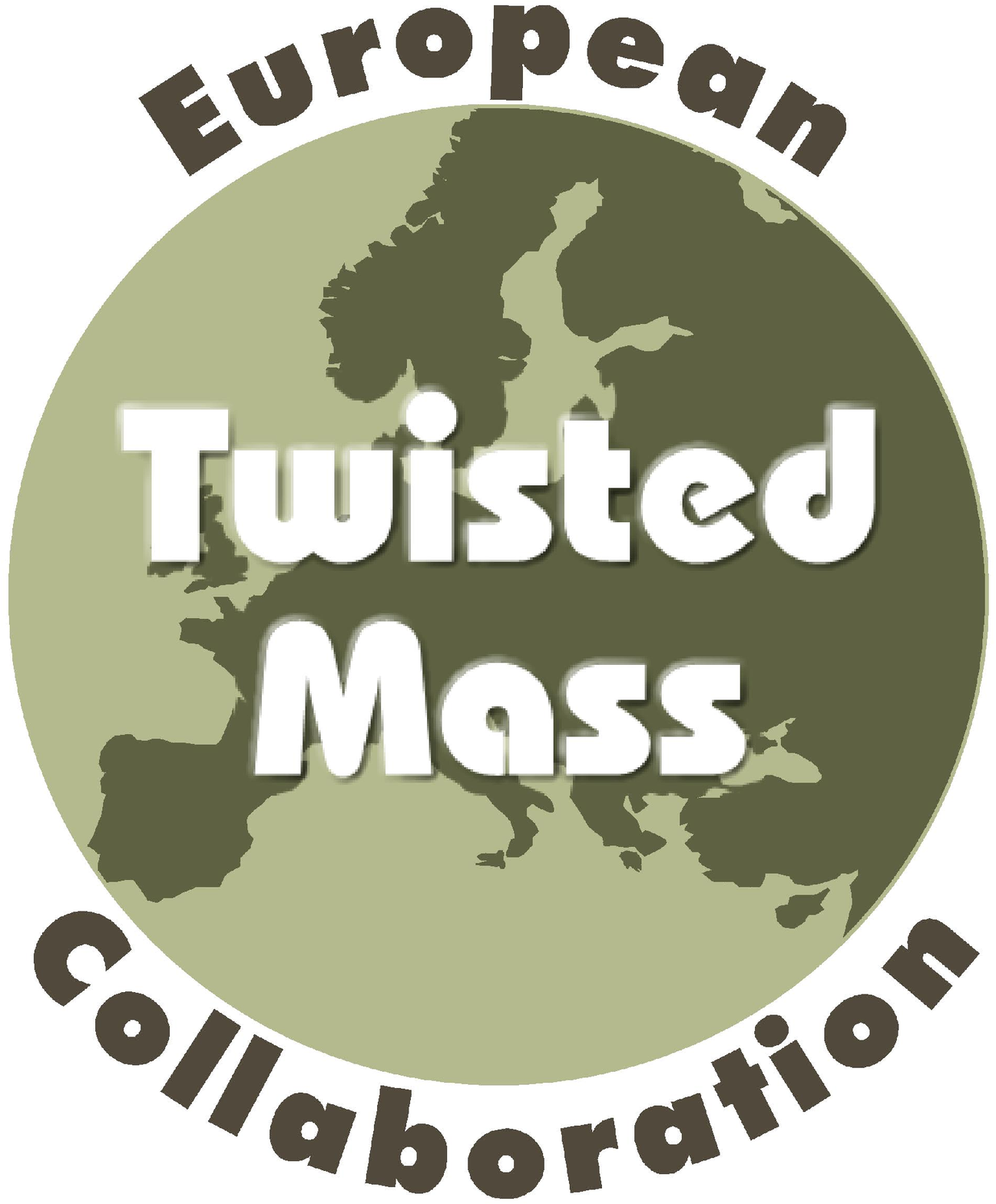}
\end{center}
}

\begin{document}
\maketitle

\section{Introduction}
\label{sec:intro}

In two recent papers \cite{Cichy:2013gja,Cichy:2013rra} we carried 
out a calculation for the chiral condensate and the topological 
susceptibility in the chiral and the continuum limit. 
In these works, we performed an  
O($a^2$) scaling towards the continuum limit. 
However, quantities such as the chiral condensate or
the topological susceptibility are related to correlators where the 
coordinates of their fields content are integrated over the whole space-time
volume~\cite{Giusti:2008vb}. This integration generates contact terms when two or more fields are
located at the same point.
The presence of contact terms can generate short-distance singularities
and when this happens, the renormalization and the discretization 
effects of these correlators need a specific
discussion. 
As we will show in this paper for the setup of maximally twisted 
mass fermions used in refs.~\cite{Cichy:2013gja,Cichy:2013rra}, even 
in the presence of these 
short distance singularities automatic O($a$)-improvement is preserved 
at maximal twist, thus justifying the strategy to 
perform an O($a^2$) scaling, as done in refs.~\cite{Cichy:2013gja,Cichy:2013rra}.

Cut-off effects in lattice correlators are described by 
the so-called Symanzik effective theory~\cite{Symanzik:1983dc,Luscher:1996sc}.
One of the basic assumptions for the validity of the Symanzik
expansion is the absence of contact terms in the lattice correlators.
These short-distance singularities alter the form of the lattice artifacts
predicted by the Symanzik effective theory.
This has been discussed already for Wilson fermions in ref.~\cite{Giusti:2008vb},
where specific O($a$) counterterms had to be added to the lattice
correlators to cancel O($a$) terms arising from the presence of short distance singularities
in the lattice correlators.

Also the property of automatic O($a$) improvement~\cite{Frezzotti:2003ni} for Wilson twisted mass 
fermions~\cite{Frezzotti:2000nk} at maximal twist relies on the validity of the Symanzik expansion 
of lattice correlators.
It is natural then to question if this property is still valid in the presence of contact terms. 
The tool to analyze the nature of the contact terms is the Operator Product Expansion
(OPE)~\cite{Wilson:1969zs}.
Using the OPE, it is possible to analyze if additional terms 
need to be added to the standard Symanzik expansion of lattice correlators.
The symmetry transformation properties of these
terms will depend on the quantity to be considered 
and the corresponding nature of the contact terms and on the lattice symmetries.

While we concentrate here on 
the example of the chiral condensate and the
topological susceptibility, we mention that a similar problem 
emerges for the vacuum polarization function needed to evaluate 
hadronic contributions to electroweak observables \cite{Renner:2012fa}, 
in particular the muon anomalous 
magnetic moment \cite{Burger:2013jya}. 
In the work here, we show which of such terms relevant for 
the chiral condensate and the topological susceptibility 
appear in this case
and we will demonstrate that the property of automatic O($a$) improvement is still preserved.
A first account of these results has been given in refs.~\cite{Cichy:2013yea,Cichy:2013zea}.
Our argumentation is similar to the one used in ref.~\cite{Shindler:2013bia}.
 
\section{Mixed action formulation and automatic O($a$) improvement}
\label{sec:auto}

To analyze the cutoff effects of the chiral condensate and the topological susceptibility,
we make use of a mixed action approach for the valence and sea quarks.
In this section, we briefly recall how the property of automatic
O($a$) improvement extends to this particular framework.
For simplicity, we consider a Wilson twisted mass (Wtm) doublet of sea quarks and 
$N_v$ Wilson twisted mass valence doublets. The extension to the 
case of $N_f=2+1+1$ Wilson twisted mass quarks~\cite{Frezzotti:2003xj} is straightforward once
the renormalized quark masses have been properly matched.

The lattice action
\be
S = S_G + S_{\rm F} + S_{\rm F,val} + S_{\rm PF}\,,
\ee
has a term for the sea quarks that reads
\be
S_{\rm F} = a^4\sum_x \chibar_s(x) \left[D_m + i\mu_s\gamma_5\tau^3\right]\chi_s(x)\,,
\ee
where
\be
D_m = \frac{1}{2}\left[\gamma_\mu\left(\nabla_\mu + \nabla_\mu^*\right) - a\nabla^*_\mu\nabla_\mu\right]+ m_0\,,
\ee
is the usual Wilson operator and $m_0$, $\mu_s$ are the bare
untwisted and twisted quark mass.
The theory contains also $N_v$ valence quark doublets with the action
\be
\label{eq:val}
S_{\rm F,val} = a^4\sum_x \sum_{v=1}^{N_v}\chibar_v(x) \left[D_m +
i\mu_v\gamma_5\tau^3\right]\chi_v(x)\,,
\ee
where $\mu_v$ denotes the valence bare twisted mass. The fermion doublets are
$\chi_v^T = (u_v,d_v)$ and $\chi_s^T = (u_s,d_s)$.

Apart from one valence quark doublet which has an associated sea quark doublet, the
additional valence doublets need appropriate pseudo-fermion fields
$\phi_v$ to cancel the valence fermionic determinant \cite{Giusti:2008vb}.
The action $S_{\rm PF}$ for the $N_v-1$ pseudo-fermion fields, i.e. complex commuting spinor
fields, is taken in the following form (in analogy to sec.~6 of ref.~\cite{Luscher:2004fu}):
\begin{equation}
 S_{\rm PF}=a^4\sum_x \sum_{v=1}^{N_v-1}\left|\left[D_m +
i\mu_v\gamma_5\right]\phi_v(x)\right|^2\,.
\end{equation} 
Note that in the above equation we use the 1-flavour twisted mass Dirac operator.

For the discussion that follows, we do not need the exact form of the gauge action $S_G$.

The long distance properties of the lattice theory close to the continuum limit
are described in terms of the Symanzik effective theory with the action
\be
S_{\rm eff} = S_0+aS_1+\ldots,
\label{eq:eff_action}
\ee
where the leading term, $S_0$, is the action of the target continuum theory
with properly renormalized parameters.
The higher order terms are linear combinations of higher-dimensional operators
\be
S_1 = \int d^4x \sum_i c_i(g_0^2) \mcO_{i}(x)\,,
\ee
where $\mcO_i(x)$ respect the symmetries of the lattice action
and we omit for simplicity the dependence on the renormalization scale.

A correlation function of products of multiplicatively renormalizable lattice fields, here denoted by
$\phi_R=Z_\phi \phi$,
at separated points $x_i$ 
\be
G(x_1,\ldots,x_n) = \langle \phi_R(x_1) \cdots \phi_R(x_n) \rangle \equiv \langle \Phi_R \rangle
\label{eq:phi}
\ee
takes the form
\be
\left\langle \Phi_R \right\rangle =  \langle \Phi_0 \rangle_0 - a \langle \Phi_0
     S_1 \rangle_0 + a \langle \Phi_1 \rangle_0 + {\rm O}(a^2)\,,
\label{eq:sym_exp1}
\ee
where 
\be
\langle \Phi_0 \rangle_0 \equiv \langle\phi_0(x_1) \cdots \cdots \phi_0(x_n) \rangle_0\,,
\ee
\be
\langle \Phi_1 \rangle_0 \equiv \sum_{k=1}^{n}\langle\phi_0(x_1) \cdots \phi_1(x_k) \cdots \phi_0(x_n) \rangle_0\,,
\ee
and $\phi_0,\phi_1$ are renormalized continuum fields. $\phi_1$ is a linear combination of 
local operators of dimension $d_\phi + 1$ that depend on the specific operator $\phi$ and 
are classified according to the lattice symmetries transformation properties of $\phi$.
The expectation values on the right hand side of eq.~\eqref{eq:sym_exp1} are to be taken in the continuum theory
with the action $S_0$.

For the sea and valence quarks, the higher-dimensional operators
contributing to $S_1$ of the Symanzik effective action are the same.
Using the equations of motion for the quark fields, a possible list of O($a$) terms
is~\cite{Luscher:1996sc,Frezzotti:2001ea}
\be
 \qquad {\mathcal O}_1^{(s,v)} = 
  i \chibar_{s,v}(x)\sigma_{\mu\nu}F_{\mu\nu}\chi_{s,v}(x)\,, \qquad 
\mcO_2^{(s,v)} = \mu_{s,v}^2\chibar_{s,v}(x)\chi_{s,v}(x)\,.
\label{eq:sym_op}
\ee
We omit from the list all the operators proportional to the untwisted quark mass. 
These terms do not contribute to the effective theory up to and including the O($a$) terms,
if we tune our lattice action to be at maximal twist, i.e. if we set the renormalized 
untwisted quark mass to vanish in the continuum limit.

\subsection{Automatic O($a$) improvement}

Automatic O($a$) improvement~\cite{Frezzotti:2003ni} is the property of Wtm that physical correlation functions 
made of multiplicatively renormalizable fields are free from O($a$) effects. 
This applies when the lattice parameters are tuned to obtain the vanishing of the renormalized untwisted
quark mass, $m_{\rm R} = 0$, in the continuum limit.

From the lattice perspective, this corresponds to setting the bare untwisted mass $m_0$ to its critical
value $\mcr$. The exact way this is achieved is not relevant for what follows, but for 
a discussion and further references on this topic see ref.~\cite{Shindler:2007vp}.

The relevant symmetries to prove automatic O($a$) improvement are the discrete chiral symmetry
\be
\mcR^{1,2}_{5} \colon
\begin{cases}
\chi_i(x) \rightarrow i \gamma_5 \tau^{1,2} \chi_i(x)\qquad i={\rm~sea,~valence} \\
\chibar_i(x) \rightarrow  \chibar_i(x) i \gamma_5 \tau^{1,2}\qquad i={\rm~sea,~valence} \\
\end{cases}
\label{eq:R512}
\ee
and the symmetry
\be
\mcD \colon
\begin{cases}
U(x;\mu) \rightarrow U^{\dagger}(-x-a\hat{\mu};\mu), \\
\chi_i(x) \rightarrow {\rm e}^{3 i \pi/2} \chi_i(-x)\qquad i={\rm~sea,~valence} \\ 
\chibar_i(x) \rightarrow  \chibar_i(-x){\rm e}^{3 i \pi/2}\qquad i={\rm~sea,~valence}. \\
\end{cases}
\ee
The equivalent transformations for continuum fields,
that with abuse of notation we indicate in the same way, are the same for the fermion
fields, whereas for the gauge fields the $\mcD$ transformation is
$A_\mu(x) \rightarrow -A_\mu(-x)$.
To include the twisted mass in the counting of the dimensions of the 
operators appearing in the lattice and continuum Lagrangian, one introduces the spurionic
symmetry
\be
\widetilde{\mathcal{D}} =  \mathcal{D}\times [\mu_i \rightarrow -\mu_i]\qquad i={\rm~sea,~valence}\,.
\ee

The lattice action is invariant under the $\mcR^{1,2}_{5} \times \widetilde{\mathcal{D}}$ 
transformation.
If the target continuum theory has a vanishing renormalized untwisted mass, $m_R=0$, it  
is invariant separately under the $\mcR^{1,2}_{5}$ and the $\widetilde{\mathcal{D}}$
transformations. 
This immediately implies that all the higher-dimensional operators in the Symanzik expansion
contributing to $S_1$ are odd under $\mcR^{1,2}_{5}$, thus they 
vanish once inserted in $\mcR^{1,2}_{5}$-even correlation functions.
The same argument applies for the higher-dimensional operators
appearing in the effective theory representations of local operators, such as axial currents
or pseudoscalar densities.
We remind that the $\mcR^{1,2}_{5}$-even correlation functions in the continuum are 
what we denote as physical correlation functions, because in the twisted basis where we are working, the 
$\mcR^{1,2}_{5}$ symmetry transformation is a physical flavor transformation.

\section{Chiral condensate}
\label{sec:chiral}

The Banks-Casher relation \cite{Banks:1979yr} connects the low lying spectrum of the 
Dirac operator with the spontaneous chiral symmetry breaking in the following way
\begin{equation}
  \label{BCrelation}
  \lim_{\lambda\rightarrow0}\lim_{\mu_s\rightarrow0}\lim_{V\rightarrow\infty} \rho(\lambda,\mu_s)=\frac{\Sigma}{\pi}\,.
\end{equation}

Eq.~(\ref{BCrelation}) relates the chiral condensate $\Sigma$ to the spectral density $\rho(\lambda,\mu_s)$. 
The method based on spectral projectors introduced in~\cite{Giusti:2008vb} 
offers a new strategy to compute spectral observables, such as the chiral condensate, in an affordable
way \cite{Cichy:2013gja,Cichy:2013rra,Engel:2014cka}. 
Moreover it allows us, via the connection to density chains, 
to compute this quantity using a representation which is free of short distance 
singularities and therefore leads to the correct continuum limit.

The integrated spectral density, i.e. the mode number $\nu(M,\mu_s)$, is defined
as the number of eigenvalues $\lambda$ of the hermitian Dirac operator
$D^\dagger D$ below a certain threshold value $M^2$. To study the
renormalization and $O(a)$ cutoff effects properties of the mode
number, it is advantageous to consider the spectral sums
$\sigma_k(\mu_v,\mu_s)$, which are directly related to the mode number through the following
expression
\begin{equation}
  \label{rel}
\sigma_k(\mu_v,\mu_s)=\frac{1}{V}\int^{\infty}_0dM\;\nu(M,\mu_s)\frac{2kM}{(M^2+\mu_v^2)^{k+1}},
\end{equation}
where $V$ is the space-time volume. To relate the mode number to a multi-local correlation function,
it is convenient to write the spectral sums $\sigma_k$ in terms
of density chain correlation functions of twisted valence quarks with mass $\mu_v$.
In terms of twisted mass density chains, the spectral sum $\sigma_3$ reads
\begin{equation}
  \label{specsum2}
   \sigma_3(\mu_v,\mu_s)=-a^{20}\sum_{x_1,\dots,x_5}\expect{P^+_{12}(x_1) P^-_{23}(x_2) P^+_{34}(x_3)
P^-_{45}(x_4) P^+_{56}(x_5) P^-_{61}(0)},
\end{equation}
where 
\be
P_{ab}^+=\chibar_a\gamma_5\tau^+\chi_b=\ubar_a\gamma_5 d_b\,,
\ee
\be
P_{ab}^-=\chibar_a\gamma_5\tau^-\chi_b=\dbar_a\gamma_5 u_b\,,
\ee
are charged
pseudoscalar densities, $\tau^\pm$ are defined in flavor
space, $\mu_v$ is the valence twisted mass and $\mu_s$ is the sea twisted mass 
that plays the role of the physical quark mass.
In this particular example, we add 6 doublets to
the theory, which is the minimum number of flavors that still
guarantees the renormalizability, as it was stated in ref.~\cite{Giusti:2008vb}.

The spectral density and therefore the mode number is directly linked to the chiral 
condensate~\cite{Giusti:2008vb}.
The representation of the mode number and the spectral density of the 
Wilson operator through density chain correlators as in
eq.~(\ref{specsum2}) allows to discuss the renormalization and improvement
properties of such quantities.
This is particularly important when computing the mode number using Wilson twisted mass
fermions at maximal twist. 
The maximal twist condition, $m_R=0$, should guarantee automatic O($a$) improvement of all physical
quantities~\cite{Frezzotti:2003ni}.
The conditional is appropriate, because density chain correlators are affected by short distance
singularities and 
the integration over the whole space-time of such singularities generates additional O($a$) terms 
that could spoil the property of automatic O($a$) improvement.
The short-distance singularities of a product of two operators can be studied 
with the operator product expansion (OPE). 

For generic values of the untwisted and twisted mass,
the Symanzik expansion for the renormalized observable 
introduced in eq.~(\ref{specsum2}) reads
\be
\sigma_{3,R}(\mu_v,\mu_s) = \sigma_{3,R}(\mu_v,\mu_s)_0 + a\sigma_{3,R}(\mu_v,\mu_s)_1 + a\sigma_{3,R}(\mu_v,\mu_s)_{\rm ct}\,,
\ee
where
\be
 \sigma_{3,R}(\mu_v,\mu_s)_0 = -\int d^4x_1 d^4x_2 d^4x_3 d^4 x_4 d^4 x_5\expect{
   P^+_{12}(x_1)P^-_{23}(x_2)P^+_{34}(x_3)P^-_{45}(x_4)P^+_{56}(x_5)P^-_{61}(0) }_0 \,,
\ee
is the continuum expectation value.
The standard terms of the Symanzik expansion are
\bea
&& \sigma_{3,R}(\mu_v,\mu_s)_1 = \int d^4x_1 d^4x_2 d^4x_3 d^4 x_4 d^4 x_5\expect{
   P^+_{12}(x_1)P^-_{23}(x_2)P^+_{34}(x_3)P^-_{45}(x_4)P^+_{56}(x_5)P^-_{61}(0) S_1  }_0 \nonumber \\
   &-& 6 c_P(g_0^2) m_v\int d^4x_1 d^4x_2 d^4x_3 d^4 x_4 d^4 x_5\expect{
   P^+_{12}(x_1)P^-_{23}(x_2)P^+_{34}(x_3)P^-_{45}(x_4)P^+_{56}(x_5)P^-_{61}(0) }_0 
\eea
where the leading O($a$) corrections to the pseudoscalar densities are
\be
(\delta P)^\pm_{ij}(x) = m_v c_P(g_0^2) P^\pm_{ij}(x)\,
\ee
and $m_v=m_0-m_c$, where $m_c$ is the critical untwisted quark mass, commonly determined through
the condition that the PCAC quark mass vanishes.

The O($a$) terms arising from the short-distance singularities, denoted by
$\sigma_{3,R}(\mu_v,\mu_s)_{\rm ct}$, get contributions from the OPE of two or more
pseudoscalar densities at a coincident space-time point. The lowest dimensional operator that appears
in the short-distance expansion (SDE) of two pseudoscalar densities on the lattice is
\be
\ubar_a(x) \gamma_5 d_b(x) \dbar_b(0) \gamma_5 u_c(0) \underset{x \rightarrow 0}{\sim}  C_{\rm PP}(x)
\ubar_a(0)u_c(0)\,,
\ee
where $C_{\rm PP}(x)\propto|x|^{-3}$. Once we sum over $x$ the
product of the
two pseudoscalar
densities, this short distance singularity will contribute a term
\bea
\sum_{x_1} & \left\langle 
\ubar_a(x_1) \gamma_5 d_b(x_1) \dbar_b(x_2) \gamma_5 u_c(x_2) \ubar_c(x_3) \gamma_5 d_d(x_3)
\dbar_d(x_4) \gamma_5 u_e(x_4) \ubar_e(x_5) \gamma_5 d_f(x_5) \dbar_f(0) \gamma_5 u_a(0)
\right\rangle\nonumber
\\
&\rightarrow a\left\langle
\ubar_a(x_2) u_c(x_2) \ubar_c(x_3) \gamma_5d_d(x_3)
\dbar_d(x_4) \gamma_5 u_e(x_4) \ubar_e(x_5) \gamma_5 d_f(x_5) \dbar_f(0) \gamma_5 u_a(0)
\right\rangle
\eea
to the Symanzik expansion.
If we now consider the lowest dimensional operator contributing to the SDE of 3 pseudoscalar densities at
the same point, we get
\be
\ubar_a(x_2) \gamma_5 d_b(x_2) \dbar_b(x_1) \gamma_5 u_c(x_1) \ubar_c(0) \gamma_5 d_d(0)
\underset{x_1,x_2 \rightarrow 0}{\sim}  
C_{PPP}(x_2,x_1)\ubar_a(0)\gamma_5d_d(0)\,,
\ee
where $C_{PPP}(x_2,x_1)\propto|x_2|^{-3}|x_1|^{-3}$. If we now sum over $x_2$ and
$x_1$, the contribution 
of the short-distance singularities to the Symanzik expansion is an O($a^2$) effect. Products of even
more 
pseudoscalar densities in the same point will give contributions of higher power of the lattice spacing.
So up to corrections of O($a^2$), the contact terms contributions to the Symanzik expansion are
\bea
\sigma_{3,R}(\mu_v,\mu_s)_{\rm ct} &=& \int d^4 x_2d^4x_3 d^4x_4 d^4x_5 
\expect{ S^{\uparrow}_{13}(x_2)P^+_{34}(x_3)P^-_{45}(x_4)P^+_{56}(x_5)P^-_{61}(0) }_0 \nonumber \\
&+&\int d^4 x_2d^4x_3 d^4x_4 d^4x_5 \expect{P^-_{23}(x_2)P^+_{34}(x_3)P^-_{45}(x_4)P^+_{56}(x_5)S^{\downarrow}_{62}(0)}_0 \nonumber \\
&+&\int d^4 x_2d^4x_3 d^4x_4 d^4x_5 \expect{P^+_{12}(x_2)S^{\downarrow}_{24}(x_3)P^-_{45}(x_4)P^+_{56}(x_5)P^-_{61}(0)}_0 \nonumber \\
&+&\int d^4 x_2d^4x_3 d^4x_4 d^4x_5 \expect{P^+_{12}(x_2)P^-_{23}(x_3)P^+_{34}(x_4)P^-_{45}(x_5) S^{\uparrow}_{51}(0)}_0 \nonumber \\
&+&\int d^4 x_2d^4x_3 d^4x_4 d^4x_5 \expect{P^+_{12}(x_2)P^-_{23}(x_3) S^{\uparrow}_{35}(x_4)P^+_{56}(x_5)P^-_{61}(0)}_0 \nonumber \\
&+&\int d^4 x_2d^4x_3 d^4x_4 d^4x_5
\expect{P^+_{12}(x_2)P^-_{23}(x_3)P^+_{34}(x_4)S^{\downarrow}_{46}(x_5)P^-_{61}(0)}_0\,,
\eea
where $S_{ac}^{\uparrow,\downarrow}=\chibar_a\frac{1}{2}(\mathbbm{1} \pm\tau^3)\chi_c$, i.e.
$S_{ac}^{\uparrow}= \ubar_a u_c\,, S_{ac}^{\downarrow}=\dbar_a d_c$.

For the discussion of the contact terms, we keep generic values for the twisted and untwisted
quark masses. To show that the contact terms $\sigma_{3,R}(\mu_v,\mu_s)_{\rm ct}$ vanish at maximal
twist, we group them and as an example we consider the two terms
\begin{eqnarray}
\label{eq:ct}
&& \int d^4 x_2d^4x_3 d^4x_4 d^4x_5 \expect{S^{\uparrow}_{13}(x_2)P^+_{34}(x_3)P^-_{45}(x_4)P^+_{56}(x_5)P^-_{61}(0)}_0+\nonumber \\
&+&\int d^4 x_2d^4x_3 d^4x_4 d^4x_5 \expect{P^-_{23}(x_2)P^+_{34}(x_3)P^-_{45}(x_4)P^+_{56}(x_5)S^{\downarrow}_{62}(0)}_0\,.
\end{eqnarray}
We can now use the integrated non-singlet axial Ward identity (WI) to rewrite eq.~\eqref{eq:ct} 
in a convenient form.
For twisted mass fermions at a generic twist angle, the WI reads 
\begin{align}
\label{WTI1}
& \int d^4 x_2d^4x_3 d^4x_4 d^4x_5 \expect{S^{\uparrow}_{13}(x_2)P^+_{34}(x_3)P^-_{45}(x_4)P^+_{56}(x_5)P^-_{61}(0)}_0+\\
&+\int d^4 x_2d^4x_3 d^4x_4 d^4x_5 \expect{P^-_{23}(x_2)P^+_{34}(x_3)P^-_{45}(x_4)P^+_{56}(x_5)S^{\downarrow}_{62}(0)}_0 \nonumber\\
&=2m_v \int d^4 x_2d^4x_3 d^4x_4 d^4x_5 \int d^4x_1 \expect{P^+_{12}(x_1) P^-_{23}(x_2)P^+_{34}(x_3)P^-_{45}(x_4)P^+_{56}(x_5)P^-_{61}(0)}_0\,. 
\nonumber
\end{align}
All the other terms stemming from the short distance singularities can be treated in an 
analogous manner. 
Thus, tuning the lattice parameters to achieve a maximal twist condition for the sea and valence quarks
guarantees that all the O($a$) terms including the non-standard ones
coming from the short-distance singularities of the correlator vanish.

For the sake of simplicity we have chosen to write a particular example for
six flavors, however, a generalization of this derivation for a
generic number of flavors is straightforward.

\section{Topological susceptibility}
\label{sec:topo}

In the continuum, the relation between the topological
charge $Q$ and the density chain correlation functions can be
established via the equation $\textrm{Tr}\{\gamma_5f(D)\}=f(0)Q$,
where $D$ is the Dirac operator and $f(\lambda)$ is any continuous function that decays rapidly enough
at infinity \cite{Luscher:2004fu}.

{With twisted mass fermions, the topological susceptibility can be defined by:
\begin{equation}
\label{chitop-tm}
\chi_{top}=\mu_{v,R}^8\; C_{4;4,R}=\frac{Z_S^2}{Z_P^2}\;\mu_{v}^8\; C_{4;4}\equiv\frac{\langle
Q^2\rangle}{V}\,,
\end{equation}
where $V$ is the space-time volume and the subscript $R$ denotes renormalized quantities,
\begin{equation}
\mu_{v,R}=Z_P^{-1}\mu_v, 
\end{equation}
\begin{equation}
C_{4;4,R}=Z_P^6 Z_S^2 C_{4;4}, 
\end{equation}
$Z_S$ and $Z_P$ are the renormalization constants of the scalar and pseudoscalar densities,
respectively,
and  
\begin{equation}
C_{4;4} = a^{28}\sum_{x_1\dots x_7}\langle S^+_{41}(x_1)P^-_{12}(x_2)P^+_{23}(x_3)P^-_{34}(x_4)
S^+_{85}(x_5)P^-_{56}(x_6)P^+_{67}(x_7)P^-_{78}(0)\rangle\,,
\end{equation}
with $S^\pm_{ij}=\chibar_{i}\tau^\pm\chi_{j}$, 
$P^\pm_{ij}=\chibar_{i}\tau^\pm\gamma_5\chi_{j}$.}
This definition of $\chi_{top}$ is interesting, because it is expressed
in terms of a correlation function of local operators, thus it can 
be used to discuss renormalization and O($a$) improvement\footnote{Note that the example that we
discuss differs from the one in ref.~\cite{Luscher:2004fu}, since we are interested in a
formula that can be evaluated with the spectral projector method and thus one that can be expressed using
the Hermitian Dirac operator $D^\dagger D$.}.
Additionally, it is directly related to the following spectral sum:
\begin{equation}
 \sigma_{k;l}(\mu) = \left\langle \Tr \left\{\gamma_5(D^\dagger D+\mu^2)^{-k}\right\}
\Tr \left\{\gamma_5(D^\dagger D+\mu^2)^{-l}\right\} \right\rangle
\end{equation} 
and hence its computation can be
carried out with the spectral projector method~\cite{Luscher:2010ik}.

In eq.~\eqref{chitop-tm}, we take $Q^2$ expressed in terms of two closed density chains -- both with 4
densities.
Note that in the case of full QCD, we could have taken one of the two density chains to contain only 2
densities -- the total of 6 densities would still guarantee the absence of
non-integrable short-distance singularities.
However, in the present case, the theory contains also pseudo-fermion fields, which allow for the
construction of flavor-singlet fields of dimension 2\footnote{Note that
this does not affect the discussion for the chiral condensate, since there are already more
than 2 densities to guarantee the absence of short-distance singularities.}.
Hence, the lowest dimensional operator appearing in
the OPE of the product of two densities in one of the density chains (with the structure
$S^+_{ab}(x)\,P^-_{ba}(0)$) would be of dimension 2 and thus the Wilson coefficient in this OPE would be
proportional to $|x|^{-4}$, leading to a logarithmic divergence upon space-time integration.
To avoid this behaviour, both density chains need to contain at least 3 densities (as done in sec.~6 of
ref.~\cite{Luscher:2004fu}).

The $\chi_{top}$ given by the above formula is
$\mathcal{R}_5^{1,2}$-even up to a charge conjugation transformation:
\be
\mathcal{C} \colon
\begin{cases}
\chi_i(x) \rightarrow C^{-1} \chibar_i(x)^T \\
\chibar_i(x) \rightarrow -\chi_i(x)^T C, \\
\end{cases}
\label{eq:C}
\ee 
where $C=i\gamma_0\gamma_2$ can be chosen.
Thus, the standard terms in the Symanzik expansion of $C_{4;4}$ vanish.
However, automatic $O(a)$ improvement can still be spoiled by contact terms.

The Symanzik expansion of $C_{4;4}$
\be
C_{4;4} = \left(C_{4;4}\right)_0 + a \left(C_{4;4}\right)_1 +a \left(\delta C_{4;4}\right)_{\rm ct} 
\ee
contains the continuum correlator $\left(C_{4;4}\right)_0$ and the standard
O($a$) terms $\left(C_{4;4}\right)_1$ coming from the higher dimensional operator
in the effective action and the effective operators.
Additional terms labeled here as $\left(\delta C_{4;4}\right)_{\rm ct}$
correspond to the O($a$) terms arising from the short distance singularities in the 
product of two densities. The product of 2 pseudoscalar densities is already discussed in the previous
section. The lowest dimensional operator appearing in the 
OPE of the product of a scalar and pseudoscalar density on the lattice is
\be
\ubar_a(x)  d_b(x) \dbar_b(0) \gamma_5 u_c(0) 
\underset{x \rightarrow 0}{\sim}  C_{SP}(x) \ubar_a(0)\gamma_5 u_c(0)\,,
\ee
where $C_{\rm SP}(x)\propto|x|^{-3}$. Once we sum over $x$ the
product of the two pseudoscalar
densities, this short distance singularity will contribute a term
\bea
\sum_{x_1} & \left\langle \ubar_a(x_1)  d_b(x_1) \dbar_b(x_2) \gamma_5 u_c(x_2) 
P^+_{cd}(x_3) P^-_{da}(x_4) 
S^+_{he}(x_5)P^-_{ef}(x_6)P^+_{fg}(x_7)P^-_{gh}(0)
\right\rangle\nonumber
\\
& \rightarrow 
a \left\langle \ubar_a(x_2)\gamma_5 u_c(x_2) 
P^+_{cd}(x_3) P^-_{da}(x_4) 
S^+_{he}(x_5)P^-_{ef}(x_6)P^+_{fg}(x_7)P^-_{gh}(0)
\right\rangle\,, 
\eea
to the Symanzik expansion (we only write the modified densities in terms of quark fields). As for the
case of the scalar condensate, contact terms
arising when 3 or more densities are at the same point lead to cut-off effects of $O(a^n)$ with $n\geq2$.
The O($a$) corrections arising from the short-distance singularities are
\begin{align}
\label{eq:ex_ct}
&\left(\delta C_{4;4}\right)_{\rm ct} = \nonumber\\
& =  c(g_0^2)  \int d^4x_2 d^4x_3
d^4x_4 d^4x_5 d^4x_6
d^4x_7\expect{P^\uparrow_{42}(x_2)P^+_{23}(x_3)P^-_{34}(x_4)S^+_{85}(x_5)P^-_{56}(x_6)P^+_{67}
(x_7)P^-_{78}(0)}_0
\nonumber\\
&+c(g_0^2)\int d^4x_1 d^4x_2 d^4x_3
d^4x_5d^4x_6
d^4x_7\expect{P^\downarrow_{31}(x_1)P^-_{12}(x_2)P^+_{23}(x_3)S^+_{85}(x_5)P^-_{56}(x_6)P^+_{67}
(x_7)P^-_{78}(0)}_0 \nonumber \\
&+c(g_0^2)\int d^4x_1d^4x_3d^4x_4d^4x_5d^4x_6
d^4x_7\expect{S^+_{41}(x_1)S^\downarrow_{13}(x_3)
  P^-_{34}(x_4)S^+_{85}(x_5)P^-_{56}(x_6)P^+_{67}
(x_7)P^-_{78}(0)}_0 \nonumber\\
&+c(g_0^2)\int d^4x_1d^4x_2d^4x_4d^4x_5d^4x_6
d^4x_7\expect{S^+_{41}(x_1) P^-_{12}(x_2) S^\uparrow_{24}(x_4)
S^+_{85}(x_5)P^-_{56}(x_6)P^+_{67}
(x_7)P^-_{78}(0)}_0\nonumber\\
&+\textrm{analogously for the 2nd density chain},
\end{align}
where $P^{\uparrow,\downarrow}_{ij}=\chibar_{i}\left(\frac{\mathbbm{1}\pm\tau^3}{2}\right)\gamma_5\chi_{j}$. 
We study now how $\left(\delta C_{4;4}\right)_{\rm ct}$ transforms under the
$\mathcal{R}_5^{1,2}$ symmetry.
Let us start considering the first two terms 
in eq.~\eqref{eq:ex_ct}. If we perform
an $\mathcal{R}_5^{1}$ transformation only for doublets labeled by $1,2,3,4$, we obtain
\begin{align}
\label{obs1}
&\expect{P^\uparrow_{42}P^+_{23}P^-_{34}S^+_{85}P^-_{56}P^+_{67}P^-_{78}}_0
+\expect{P^\downarrow_{31}P^-_{12}P^+_{23}S^+_{85}P^-_{56}P^+_{67}P^-_{78}}_0\nonumber\\
&\xrightarrow{\mathcal{R}_5^1}
-\expect{P^\downarrow_{42}P^-_{23}P^+_{34}S^+_{85}P^-_{56}P^+_{67}P^-_{78}}_0
-\expect{P^\uparrow_{31}P^+_{12}P^-_{23}S^+_{85}P^-_{56}P^+_{67}P^-_{78}}_0\,.
\end{align}
Up to a relabeling of flavors ($4\rightarrow3$, $3\rightarrow2$, $2\rightarrow1$ in the
first term and $1\rightarrow2$, $2\rightarrow3$, $3\rightarrow4$ in the
second one), this linear combination is odd under $\mathcal{R}_5^{1,2}$, 
i.e. it vanishes for twisted mass fermions at maximal twist.
For the third and the fourth term in eq.~\eqref{eq:ex_ct},
after the $\mathcal{R}_5^{1}$ transformation on the doublets $1$ to $4$, we obtain
\begin{align}
\label{obs1}
&\expect{S_{41}^+S^\downarrow_{13}P^-_{34}S^+_{85}P^-_{56}P^+_{67}P^-_{78}}_0
+\expect{S_{41}^+P^-_{12}S^\uparrow_{24}S^+_{85}P^-_{56}P^+_{67}P^-_{78}}_0\nonumber\\
&\xrightarrow{\mathcal{R}_5^1}
-\expect{S_{41}^-S^\uparrow_{13}P^+_{34}S^+_{85}P^-_{56}P^+_{67}P^-_{78}}_0
-\expect{S_{41}^-P^+_{12}S^\downarrow_{24}S^+_{85}P^-_{56}P^+_{67}P^-_{78}}_0\nonumber\\
&\xrightarrow{\mathcal{C}}-\expect{S_{14}^+S^\uparrow_{31}P^-_{43}S^+_{85}P^-_{56}P^+_{67}P^-_{78}}_0
-\expect{S_{14}^+P^-_{21}S^\downarrow_{42}S^+_{85}P^-_{56}P^+_{67}P^-_{78}}_0\nonumber\\
&\xrightarrow{\rm relabel}
-\expect{S_{41}^+P^-_{12}S^\uparrow_{24}S^+_{85}P^-_{56}P^+_{67}P^-_{78}}_0
-\expect{S_{41}^+S^\downarrow_{13}P^+_{34}S^+_{85}P^-_{56}P^+_{67}P^-_{78}}_0,
\end{align}
where the relabeling of the doublets is $1\leftrightarrow4$, $2\leftrightarrow3$.
Thus, also the sum of the third and fourth terms in eq.~\eqref{eq:ex_ct}
is odd under the symmetries of the action and thus vanishes.
The same procedure can be used also for the second closed density chain of
eq.~\eqref{eq:ex_ct},
applying the $\mathcal{R}_5^{1}$ transformation only to doublets labeled by $5-8$.
Moreover, this proof holds also in the general case -- for any
density chain that can be written in terms of $D^\dagger D$ (i.e. containing an even (and not
smaller than 4) number of
pseudoscalar and scalar densities in each density chain).

\section{Concluding remarks}
\label{sec:conclu}

When using density chain correlators to compute 
the chiral condensate and the topological susceptibility as suggested in 
ref.~\cite{Giusti:2008vb}, short distance singularities appear. 
Thus, the influence of resulting contact terms 
needs to be analyzed. In particular, it is a priori unclear, 
whether the property of automatic O($a$) improvement is 
preserved for maximally Wilson twisted mass fermions in the presence of such terms. 

Contact terms arise in lattice correlators when two or more (pseudo)scalar densities are at the same
space-time point and generate short-distance singularities that can spoil this 
automatic O($a$) improvement, i.e.
introduce O($a$) cut-off effects in physical correlators.
Working in the framework of the Operator Product Expansion and using 
the symmetries of our setup, we have shown that the additional terms in
the Symanzik expansion that arise due to contact terms vanish at maximal twist.
Thus, automatic O($a$) improvement is preserved, justifying the O($a^2$) 
continuum limit scaling analysis of refs.~\cite{Cichy:2013gja,Cichy:2013rra}. 

We remark that our discussion holds also in the general case -- for any density chain that can be written
in terms of $D^\dagger D$ (i.e. containing an even number of pseudoscalar and scalar densities). 
For a discussion concerning the automatic O($a$) improvement of the hadronic vacuum polarization function, we refer to Ref.~\cite{Hotzel}.

\section*{Acknowledgments}
We would like to thank the referee of this paper for the very constructive reports which helped us to find and correct a flaw in the argument of the O(a)-improvement of the topological susceptibility.
For the analysis carried through in this paper discussions with Andrea Shindler have been instrumental
and we gratefully acknowledge his significant help and the development of the strategy for the proof of
O($a$) improvement presented in this paper. 
We also thank Vincent Drach, Gregorio Herdoiza, Alberto Ramos and Giancarlo Rossi for useful discussions
and Marcus Petschlies for carefully reading the manuscript.
This work has been supported in part by the Foundation for Polish Science fellowship ``Kolumb''
and by the DFG Sonderforschungsbereich/Transregio SFB/TR9.

\vspace{0.4cm}

\bibliography{sd_auto}  
\bibliographystyle{jhep}

\end{document}